\documentclass[prc,twocolumn,tightenlines,nofootinbib,superscriptaddress]{revtex4-1}
\usepackage{tipa}
\usepackage{mathrsfs}
\usepackage{amsmath}
\usepackage{amssymb}
\usepackage{graphicx}
\usepackage{color}
\usepackage{tikz}
\usepackage[T1]{fontenc}
\usepackage[english]{babel}
\usepackage[utf8]{inputenc}
\usepackage{soul}
\newcommand{\be}{\begin{eqnarray}}
\newcommand{\ee}{\end{eqnarray}}

 \newcommand{\gsim}{\mathrel{\hbox{\rlap{\lower.55ex \hbox {$\sim$}}
                   \kern-.3em \raise.4ex \hbox{$>$}}}}
\newcommand{\lsim}{\mathrel{\hbox{\rlap{\lower.55ex \hbox {$\sim$}}
                   \kern-.3em \raise.4ex \hbox{$<$}}}}

\begin{document}

\title{  Gluon Bremsstrahlung in Relativistic Heavy Ion Collisions }

\author{Mawande Lushozi}
\email{mlushozi@uw.edu}
\affiliation{Institute for Nuclear
	Theory, University of Washington, Box 351550, Seattle, WA, 98195, USA  }

\author{Larry D. McLerran}
\email{mclerran@me.com}
\affiliation{Institute for Nuclear
Theory, University of Washington, Box 351550, Seattle, WA, 98195, USA  }

\author{Michal Praszalowicz}
\email{michal@if.uj.edu.pl}
\affiliation{Institute of Theoretical Physics, Jagiellonian University, S.{\L}ojasiewicza 11, 30-348 Krak\'ow, Poland }

\author{Gongming Yu}
\email{ygmanan@uw.edu}
\affiliation{Institute for Nuclear Theory, University of Washington, Box 351550, Seattle, WA, 98195, USA }
\affiliation{CAS Key Laboratory of High Precision Nuclear Spectroscopy and Center for Nuclear Matter Science, Institute of Modern Physics, Chinese Academy of Sciences, Lanzhou 730000, China}

\begin{abstract}
We study the process $qq\rightarrow qqg$ at lowest order in QCD perturbation theory to understand gluon radiation in the fragmentation region of relativistic heavy-ion collisions. We arrive at a formula for gluon multiplicity that interpolates between  $\sim 1/k_{\bot}^2$ behavior at low $k_{\bot}$, to $\sim 1/k_{\bot}^4$ at large $k_{\bot}$.


\end{abstract}

\maketitle


\let\clearpage\relax
\section{Introduction}

Understanding gluon distributions in the fragmentation region of ultra-relativistic heavy ion collisions 
 \cite{AKM1980fragmention,M2018fragmention} is interesting because it gives insight into the behavior of matter at high baryon density in the presence of strong gluon fields.  The ultimate application we envisage for this work is to understand the initial conditions for hydrodynamic or transport for the valence quarks and associated radiation produced in the fragmentation region of heavy ion collisions.  Such matter may form a quark-gluon plasma that is distinctively different from that found in the central region.  There may be a finite ratio of baryon number chemical potential to temperature.  The earliest estimates suggested that energy densities and times scales are sufficient for a formation of a quark gluon plasma \cite{AKM1980fragmention}.  What is required is to update these old estimates in light of what we have understood about the Color Glass Condensate and the coherence of particle production.

In this work, we will be concerned with gluon radiation in the fragmentation region of the target nucleus. 
 Even if
 the colliding nuclei are of the same size, one faces an asymmetry between the saturation scales of the target and projectile ($Q_s^{\text{targ}}\ll Q_s^{\text{proj}}$), an asymmetry which is enhanced if the projectile nucleus is larger. This is because the saturation scale of a nucleus (or hadron) is proportional to the gluon rapidity density \cite{GLR1983Saturation,M1990Saturation}, $dN/dy$, which grows like an exponential in the rapidity-difference $\tau\equiv\ln 1/x=y_{\text{nucl}}-y$ \cite{KLF1977BFKL,BL1976BFKL}. In the fragmentation region of the target nucleus, this rapidity-difference is by definition very small; but for the projectile, it is large, and hence the respective gluon densities are very different. For the ultrarelativistic case, 
  which we are interested in, not only may the gluon fields be treated classically -- as is typically the case in saturation physics, \cite{mclerran1994computing} -- but the asymmetry of this problem allows one to solve classical Yang-Mills  equations \cite{GV2006YangMills,W1970YangMills} by treating projectile as a strong background field $A^{\mu}$, while the target field, $\delta A^\mu$, is taken to first order since it is much weaker. This asymmetry of saturation scales is not unique to the case of rapidities far from the central region, in fact it has been exploited to calculate gluon radiation in the central region of collisions involving particles with 
 different sizes (p-A for example) \cite{dumitru2002protons}. Many features of the results in that scenario
 are shared by a calculation in the fragmentation region.

In  previous works, Kajantie, McLerran, and Paatelainen \cite{kajantie2019gluon,Kajantie:2019nse}, proceeding in this spirit 
of classical Yang-Mills,  took steps towards calculating small-$k_{\bot}$ gluon radiation in the fragmentation region of nucleus-nucleus collisions. 
We will now give a very brief overview of their key results, highlighting the issues we wish to address.

In the computations of Kajantie, McLerran and Paatelainen, for a nucleus-nucleus collision, there are two sources of coherence.  
The first and easiest to treat arises from the high energy projectile that strikes a target nucleus.  This projectile may be teated as a source 
of color field, and this is straightforward to treat by methods developed for the central region.  The difficult part of the computation was 
to treat the radiation associated with coupling to sources in the fragmentation region.  It was possible to do this insofar as the gluon fields 
are treated in a no recoil approximation.  This meant that there was not a complete treatment of the high transverse momentum tail of the 
radiation distribution.  For this high momentum tail, the gluon production is no longer coherent and can be treated to first order in 
the strength of the projectile and the target sources, but the recoil can be treated to all orders.  Including recoil will complete the computation 
of such radation.

Specifically, the problem considered in \cite{kajantie2019gluon,Kajantie:2019nse} is that of gluon radiation produced when 
a sheet of colored glass interacts with a classical particle that has 
an associated color-charge vector $\mathbf{T}$. One then finds two sources of radiation, the first is ED-like bremsstrahlung of a charged particle 
getting a momentum kick from $p$ to $p'$; the second, which they term the bulk contribution, is from interaction of  a gluon emitted by 
the target quark with the sheet of colored glass, with a term with no reinteraction subtracted. The former  is calculated from the following 
radiation current,
\begin{align}
J^\mu(k)&=iT\left(\frac{p^\mu}{p\cdot k}-\frac{p'^{\mu}}{p'\cdot k} \right)\,\,,
\end{align}
with the resulting gluon distribution \cite{kajantie2019gluon}
\begin{align}\!\!\!\!\!16\pi^3|\mathbf{k}|\frac{d N}{d^3k} &=J^{\mu}(k)J_{\mu}(-k)\nonumber \\
	&=T^2\!\left[m^2\!\left(\!\frac{1}{p\cdot k}\!-\!\frac{1}{p'\cdot k}\right)^{\!\!2}\!\!\!-\!\frac{p'^2_{_T}}{(p\cdot k)(p'\cdot k)} \!\right],
\end{align}
where $k$ is the gluon's momentum and $p'_{_T}$ is the transverse-momentum kick of the charged particle.
The above result falls off like $\sim 1/k_{\bot}^2$ no matter the value of $k_{\bot}$. The bulk contribution was calculated numerically for general values of ${k_{\bot}}$, and analytically in the large-$k_{\bot}$ limit with the resulting fall off $\sim \log (k_{\bot})/k_{\bot}^4$.  To understand why this result is troublesome
let's recount some findings from Ref.\,\cite{dumitru2002protons}, which, as discussed above, is analogous to the study of the fragmentation region. In the region\footnote{$Q_s^{(2)}$ and $Q_s^{(1)}$ are respectively the saturation momenta of the large nucleus and the proton. In the case of the fragmentation region, these correspond to the saturation momenta of the projectile and target nucleus respectively.} $Q_s^{(2)}>k_{\bot}>Q^{(1)}_{s}$, the gluon distribution has a $\sim 1/k_{\bot}^2$ behavior and for $k_{\bot} >Q_s^{(2)}$ it changes to $\sim 1/k_{\bot}^4$. Also note, in region $k_{\bot}> Q_s^{(2)}$, the fields of both particles are weak enough to permit a perturbative calculation. Hence as far as one can calculate, the bulk contribution is correct and  interpolates between the $\sim 1/k_{\bot}^2$ behavior at low $k_{\bot}$ to the $\sim 1/k_{\bot}^4$ fall-off at large transverse momentum. This result is remarkable, considering that this calculation is non perturbative and classical. However, the ED-like radiation shown above is not correct, at least for large transverse momentum $k_{\bot}>Q^{\text{\tiny{proj}}}_s$. Indeed the calculation as done in Ref. \cite{kajantie2019gluon,Kajantie:2019nse} was only meant to be valid for small $k_{\bot}$, i.e when the gluon does not carry away a significant fraction of the quark's momentum.

In the present work, our aim is to understand how this formula may be remedied and produce one that has the correct behavior for all values of $k_{\bot}$. Since we want to deal with large transverse momentum we can turn to a simpler but related perturbative problem. We calculate gluon radiation from quark-quark scattering to lowest order in QCD perturbation theory, where we consider one quark to be at rest and the other ultrarelativistic.

This paper is organized as follows. In Section II we derive the multiplicity distribution of gluons perturbatively using the $qq\rightarrow qqg$ process. Section III is a discussion of the results,  we study the gluon multiplicty distribution in various kinematic limits with a special focus on the fragmentation region. Section IV is the conclusion.

\begin{figure}
		\includegraphics[width=.45\textwidth]{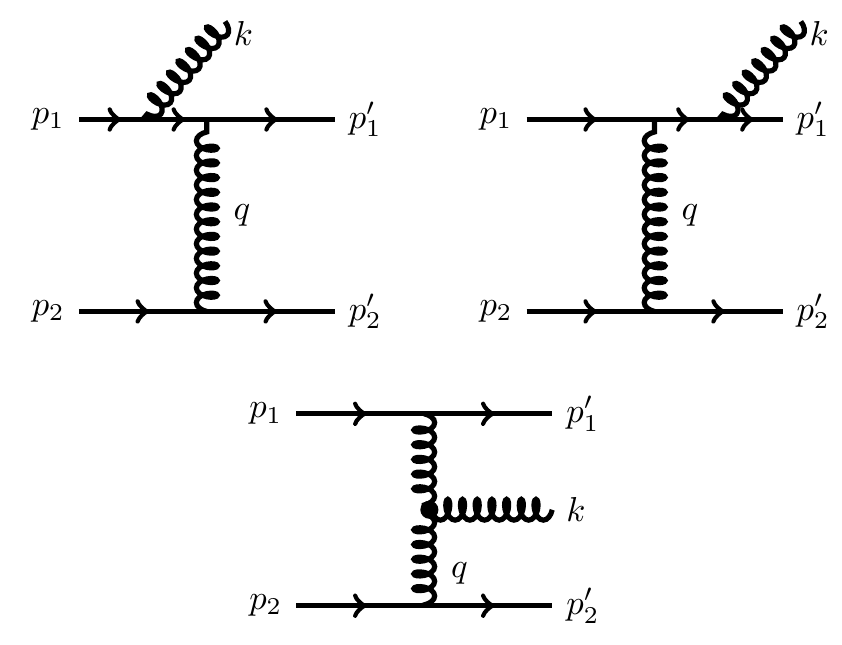}
	\caption{The lowest order tree-level diagrams for gluon bremsstrahlung.}\label{bremst}
\end{figure}

\section{Gluon Bremsstrahlung}

 We calculate gluon bremsstrahlung perturbatively using the process $qq\rightarrow qqg$  (FIG.\,\ref{bremst}). Both beam and target could move on the light cone, the beam fragmentation region can be studied in the forward limit \cite{GSV2006ForawardLimit,IMS2006ForawardLimit}, but in our calculation we just consider the target to be at rest. This treatment is similar to that of Gunion and Bertsch \cite{gunion1982hadronization,L1975,N1976}, and it is also just the Lipatov vertex \cite{KLF1976LipatovVertex} at high energy. We use the following kinematics (in Light-Cone coordinates)
\begin{eqnarray}\label{y3}
\!\!\!\!\!\!\!\!&&p^{\mu}_{1}=(M,M,\vec{0}),\,\,p^{\mu}_{2}=(0,P,\vec{0}),\,\,k^{\mu}=(xM,\frac{\vec{k}_{\bot}^{2}}{2xM},\vec{k}_{\bot}),\nonumber\\[1mm]
&&p'^{\mu}_{1}=\left((1-x)M,M+\frac{xM}{1-x}+\frac{(\vec{q}_{\bot}-\vec{k}_{\bot})^2}{2M(1-x)},\vec{q}_{\bot}-\vec{k}_{\bot} \right),\nonumber\\[1mm]
&&p'^{\mu}_{2}=\left(0,P-\frac{k^2_T }{2Mx} -\frac{xM}{1-x}-\frac{\left(\vec{q}_{\bot}-\vec{k}_{\bot}\right)^2}{2M(1-x)},-\vec{q}_\bot\right)\!\!,
\end{eqnarray}
where $p^{\mu}_{1}$ is the initial momentum of the quark at rest with $M=m/\sqrt{2}$ (where $m$ is the quark mass), $p^{\mu}_{2}$ is initial momentum of the incident quark with energy $P$, $k^{\mu}$ is the momentum of the radiated gluon, and $x$ is the fractional light-cone momentum carried by the radiated gluon. The momentum $p'^{\mu}_{1}$ and $p'^{\mu}_{2}$ are the final momentum of the two quarks, respectively. The momentum transfer can be written as
\begin{eqnarray}\label{y4}
q^{\mu}=\left(0,\frac{k^2_T }{2Mx} +\frac{xM}{1-x}+\frac{\left(\vec{q}_{\bot}-\vec{k}_{\bot}\right)^2}{2M(1-x)},\vec{q}_\bot\right),
\end{eqnarray}
where we have dropped terms of order $\sim 1/P$.

We calculate in the Light-Cone gauge, and hence the gluon propagator takes the following form
\begin{eqnarray}\label{y5}
\!\!\!\!\!\!\!iS_{B}^{\mu\nu}\!(q)\!=\!\frac{-i\delta_{ab}}{q^2}\!\left(\!g^{\mu\nu}\!-\!\frac{q^{\mu}n^{\nu}\!+\!n^\mu q^\nu}{n\cdot q}\right),
\end{eqnarray}
where $n=(0,1,\vec{0})$. With this choice, diagrams with gluon emissions from the bottom quark line do not contribute to the amplitude-squared, and hence it suffices to consider only the three diagrams in FIG.\,\ref{bremst}. 

 We define gluon multiplicity distribution in the following way. First we calculate the cross-section for gluon bremsstrahlung:

\begin{equation}
\frac{d\sigma }{dyd^{2}k_{\bot }}=\frac{1}{(2\pi )^{5}}\frac{1}{\text{flux}}%
\frac{1}{8PM}\int \left|\mathcal{M}_{(a+b+c)}\right|^{2}\frac{1}{1-x}%
d^{2}q_{\bot }.  \label{xs}
\end{equation}
Factor $1/(1-x)$ in (\ref{xs}) comes from the phase-space integration
\begin{equation}
\int dq^{-}\,\delta _{+}(p_{1}^{\prime \,2}-m^{2}) 
\label{eq:delta}
\end{equation}
where
\begin{align}
p_{1}^{\prime \,2}-m^{2}= &2(q^{+}+M-k^{+})\, q^{-} 
-(q_{\bot }-k_{\bot})^{2}\notag \\
-&2M(k^{+}-q^{+})-k_{\bot }^{2}\frac{M-k^{+}+q^{+}}{k^{+}}.
\label{p1p2}
\end{align}
From the on-shell condition $(p_{2}-q)^{2}=0$ we get that $q^+\sim1/P$, and the Jacobian of the delta function in (\ref{eq:delta})
is $1/(2M(1-x))$ in the linit $P\rightarrow \infty$.

On the other hand the Born cross-section for quark-quark scattering {\em without} gluon radiation reads:
\begin{equation}
\frac{d\sigma _{\text{Born}}}{d^{2}q_{\bot }}=\frac{1}{(2\pi )^{2}}\frac{1}{%
\text{flux}}\frac{1}{4PM}\left\vert \mathcal{M}_{\text{Born}}\right\vert
^{2}.
\label{eq:Bornxs}
\end{equation}
where the {\em flux} factor is the same as in (\ref{xs}), and
\begin{equation}
\left\vert \mathcal{M}_{\text{Born}}\right\vert ^{2}=\frac{C_{F}}{N_{c}}%
\frac{8g^{2}P^{2}M^{2}}{q_{\bot }^{2}}.
\label{eq:Bornxs1}
\end{equation}
We define gluon multiplicity distribution from the convolution:
\begin{equation}
\frac{d\sigma }{d^{2}k_{\bot }dy} =\int d^{2}q_{\bot }\frac{d\sigma _{\text{Born}}}{d^{2}q_{\bot }}
\frac{dN}{d^{2}k_{\bot }dy}
\label{eq:gluondef}
\end{equation}

The amplitude-squared is the sum of six terms
\begin{figure}
	\centering
	\includegraphics[width=.45\textwidth]{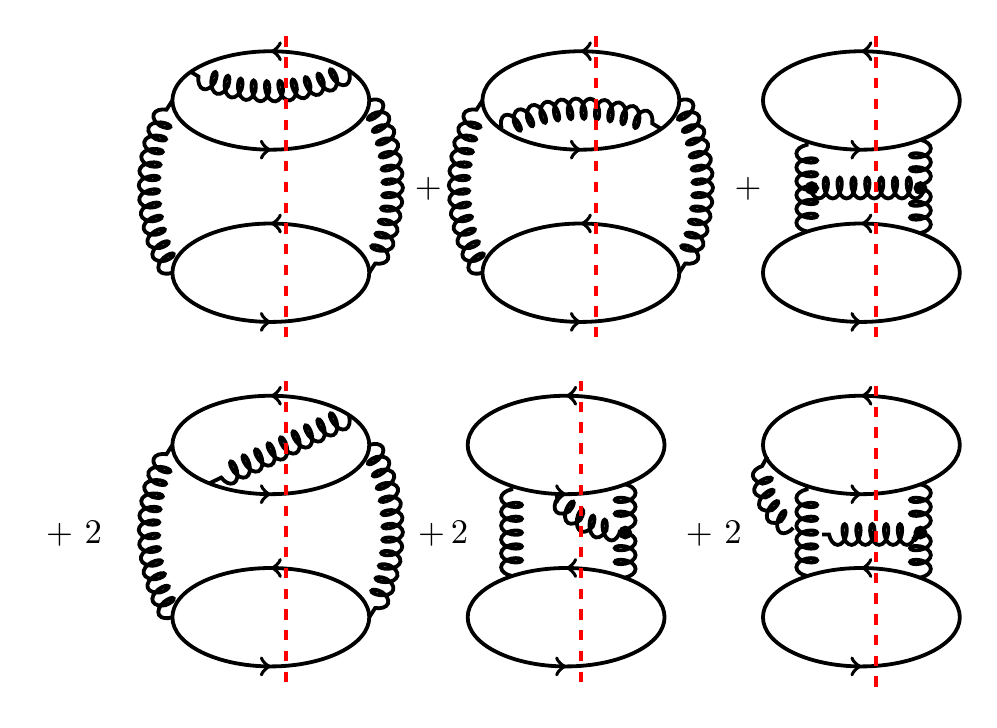}
	\caption{\label{ampsq} The six pieces of the amplitude-squared.}
\end{figure}
\begin{align}\left|\mathcal{M}_{(a+b+c)}\right|^{2}&=\left|\mathcal{M}_a\right|^2+\left|\mathcal{M}_b\right|^2+\left|\mathcal{M}_c\right|^2+2\mathcal{M}_a\mathcal{M}_b^{\ast}\nonumber\\
	&+2\mathcal{M}_b\mathcal{M}_c^{\ast}+2\mathcal{M}_a\mathcal{M}_c^{\ast}\,\,,
\end{align}
which are respectively displayed in diagramatic form in FIG.\,\ref{ampsq}. We get the following results for each of these terms
\begin{align}\left|\mathcal{M}_a\right|^2&= 2K\frac{C_F^2}{N_c}\left[\frac{x^2}{D_A}+\frac{4x^2(x-1)}{D_A^2}\right], \\
	\left|\mathcal{M}_b\right|^2&= 2K\frac{C_F^2}{N_c}\left[\frac{x^2}{D_B}+\frac{4x^2(x-1)M^2}{D_B^2}\right],\\
	\left|\mathcal{M}_c\right|^2&= K\frac{C_AC_F}{N_c}\left[\frac{2(x^2-2x+2)}{D_C}+\frac{8x^2(x-1)}{D_C^2}M^2\right],\\
\mathcal{M}_a\mathcal{M}_b^{\ast}&=-K\frac{C_F}{2N_c^2}\left[\frac{-x^2}{D_B}+\frac{x^2(x^2-2x+2)}{D_AD_B}q_T^2\right. \nonumber\\
	&\left.-\frac{x^2}{D_A}+\frac{8x^2(1-x)}{D_AD_B}M^2\right]\label{Mab}, \\
\mathcal{M}_a\mathcal{M}_c^{\ast}&=K\frac{C_AC_F}{2N_c}\left[\frac{-x^2}{D_A}+\frac{x^2-2x+2}{D_AD_C}q_{\bot}^2\right.\nonumber\\
	&\left.-\frac{x^2-2x+2}{D_C}+\frac{8x^2(1-x)}{D_AD_C}M^2 \right],\\
\mathcal{M}_b\mathcal{M}_c^{\ast}&=	K\frac{C_AC_F}{2N_c}\left[\frac{-x^2}{D_B} +\frac{(x-1)^2(x^2-2x+2)}{D_BD_C}q_{\bot}^2 \right.\nonumber\\
&\left.-\frac{x^2-2x+2}{D_C}+\frac{8x^2(1-x)}{D_BD_C}M^2\right],
\end{align}
where $D_A=k_{\bot}^2+2x^2M^2$, $D_B=\left(\vec{k}_{\bot}-x\vec{q}_{\bot}\right)^2+2x^2M^2$, $D_C=\left(\vec{k}_{\bot}-\vec{q}_{\bot}\right)^2+2x^2M^2$, ${C_A=N_{c}=3}$, $C_{F}=\frac{N_{c}^{2}-1}{2N_{c}}$, and $K=8g_s^6(1-x)M^2P^2/q_{\bot}^4$. We have only kept terms that are proportional to $P^2$, which turns out to be the leading power in the large momentum $P$.

 If we ignore Eq.\eqref{Mab}, then the terms above separate into two sets according to their color factors: either a term is proportional to $C_F$ or it is proportional to $C_F^2/N_c$. Taking alook at Eq.\eqref{Mab}, we notice that it comes with the following color factor
\begin{align}
-\frac{C_F}{2N_c^2}&= \frac{C_F^2}{N_c}-\frac{C_AC_F}{2N_c}\,\,,
\end{align}
and so we see that it contributes to both sets. In fact it contributes precisely so as to cancel out all the terms that would go as $\sim 1/k_{\bot}^2$ at large $k_{\bot}$, leaving only those that go as $\sim 1/k_{\bot}^4$. The full amplitude-squared can then be expressed as:

\begin{eqnarray}\label{y6}
&&\left|\mathcal{M}_{(a+b+c)}\right|^{2}
	=2g^2_s\, \frac{C_F}{N_c}\,\frac{8 g_s^4 M^2P^2}{q_{\bot}^4} (1-x) \\[1mm]
&&\times\bigg\lbrace C_F x^2\bigg[\frac{(x^2\!-\!2x\!+\!2)}{D_A D_B}q_{\bot}^2\!+\!4(x-1)M^2\bigg(\frac{1}{D_A}-\frac{1}{D_B}\bigg)^2 \bigg]\nonumber\\[1mm]
&&+\frac{C_A}{2}\bigg[(x^2\!-\!2x\!+\!2)q_{\bot}^2\bigg(-\frac{x^2}{D_{A}D_{B}}+\frac{1}{D_{A}D_{C}}+\frac{(x-1)^2}{D_{B}D_{c}}\bigg)\nonumber\\[1mm]
&&+8x^{2}(1\!-\!x)M^2\!\bigg(\!\!-\!\frac{1}{D_{\!A}D_{\!B}}\!+\!\frac{1}{D_{\!A}D_{\!C}}\!+\!\frac{1}{D_{\!B}D_{\!C}}\!-\!\frac{1}{D_{\!C}^2}\bigg)\!\bigg]\!\bigg\}\nonumber .
\end{eqnarray}


In the first line of Eq.~(\ref{y6}) we recognize the Born amplitude squared of Eq.~(\ref{eq:Bornxs}).
By comparing (\ref{y6}) with (\ref{eq:gluondef}) we arrive at our final result for the gluon distribution:

\begin{eqnarray}\label{y8}
&&\frac{dN}{d^{2}k_{\bot }\, dy}(\vec{q}_{\bot })= \frac{\alpha _{s}}{2\pi ^{2}} \\
&\times&\bigg\lbrace{C_{F}x^{2}}\bigg[\frac{%
(x^{2}\!-\!2x\!+\!2)}{D_{A}D_{B}}q_{\bot }^{2}\!+\!4(x-1)M^{2}\bigg(\frac{1}{%
D_{A}}-\frac{1}{D_{B}}\bigg)^{2~}\bigg]  \notag \\[1mm]
&+&\frac{C_{A}}{2}\bigg[(x^{2}\!-\!2x\!+\!2)q_{\bot }^{2}\bigg(-\frac{x^{2}}{%
D_{A}D_{B}}+\frac{1}{D_{A}D_{C}}+\frac{(x-1)^{2}}{D_{B}D_{c}}\bigg)  \notag
\\[1mm]
&+&8x^{2}(1\!-\!x)M^{2}\!\bigg(\!\!-\!\frac{1}{D_{\!A}D_{\!B}}\!+\!\frac{1}{%
D_{\!A}D_{\!C}}\!+\!\frac{1}{D_{\!B}D_{\!C}}\!-\!\frac{1}{D_{\!C}^{2}}\bigg)%
\!\bigg]\!\bigg\}.\notag
\end{eqnarray}%

\section{Discussion}

To start analyzing the bremstrahlung probability in Eq.\eqref{y8}, we will
 assume that $M^{2}\ll k_{\bot}^{2},q_{\bot}^{2}$, which allows one to
neglect $M^{2}$ in the numerators and in $D_A$, but not in $D_{B,C}$ where
$M^{2}$ regularizes singularities when $k_{\bot}\sim q_{\bot}$ or $k_{\bot
}\sim xq_{\bot}$,
\begin{align}
\frac{dN}{d^{2}k_{\bot}\, dy}&\sim\alpha_{\text{s}}q_{\bot}^{2}(x^{2}%
-2x+2)\left\{  2C_{F}\frac{\,x^{2}}{D_B D_A}\right.\nonumber\\
&\left.+C_{A}\left(  \frac{(1-x)^{2}%
}{D_B D_C }+\frac{1}{D_AD_C }-\frac{x^{2}}{D_B D_A}\right)  \right\}
.
\end{align}
A few remarks are in order here: the part proportional to $C_{F}$ can be viewed as
 bremsstrahlung from a fermion line (including the interference), whereas
terms proportional to $C_{A}$ involve a 3-gluon vertex and a $C_{A}$ part of
fermion interference ($x^{2}/\left(  D_B D_A\right)  $). Importantly, a
term from the 3-gluon vertex squared that should naively be proportional to
$1/D_C ^{2}$ cancels out (except for a term proportional to $M^{2}$ that we
neglected). This is why the full result is proportional to $(x^2-2x+2)$, which is connected to the DGLAP probability,%
\begin{equation}
P_{q\rightarrow g}=C_{F}\frac{\alpha_{\text{s}}}{2\pi}\frac{1+(1-x)^{2}}{x}.
\end{equation}
Note also that in the
large $N_{c}$ limit $2C_{F}\rightarrow N_{c}=C_{A}$, and both terms have the
same color factor.

Let us first check limits coming from $x=1$ or $x=0$ (note that in these
limits the mass terms that we neglected vanish):

\begin{itemize}
	\item  For $x=0$ we have $D_B =D_A=k_{\bot}^{2}$ and ${D_C =(\vec{k}_{\bot
	}-\vec{q}_{\bot})^{2}}$,%
	\begin{equation}
	\left.  \frac{dN}{d^{2}k_{\bot} dy}\right\vert _{x=0}\sim4\alpha_{\text{s}%
	}C_{A}\frac{q_{\bot}^{2}}{k_{\bot}^{2}(\vec{k}_{\bot}-\vec{q}_{\bot})^{2}%
	}.\label{xeq0}%
	\end{equation}
	This is the Bertsch-Gunion (BG) formula \cite{gunion1982hadronization}. Note that it comes entirely from the
	interference term.
	
	\item For $x=1$ we have $D_B =D_C =(\vec{k}_{\bot}-\vec{q}_{\bot}%
	)^{2}+2M^{2}$ and $D_A=k_{\bot}^{2}$,
	\begin{equation}
	\left.  \frac{dN}{d^{2}k_{\bot} dy}\right\vert _{x=1}\sim2\alpha_{\text{s}%
	}C_{F}\frac{q_{\bot}^{2}\,}{k_{\bot}^{2}\left(  (\vec{k}_{\bot}-\vec{q}_{\bot
		})^{2}+2M^{2}\right)  }.\label{xeq1}%
	\end{equation}
	This looks like the BG formula, but with a different color factor. The numerical
	coefficient in front is $2$ rather than 4 due to a different value of
	$(x^{2}-2x+2)$ at $x=1$ and 0.
\end{itemize}

Now we shall investigate three limits in $k_{\bot}$:

\begin{itemize}
	\item For $k_{\bot}\ll xq_{\bot}$ we have $D_B =x^{2}q_{\bot}^{2}%
	,D_A=k_{\bot}^{2}$ and $D_C =q_{\bot}^{2}$,%
	\begin{equation}
	\frac{dN}{d^{2}k_{\bot} dy}\sim\alpha_{\text{s}}(x^{2}-2x+2)\left\{
	2C_{F}\frac{\,1}{k_{\bot}^{2}}+C_{A}\frac{(1-x)^{2}}{x^{2}q_{\bot}^{2}%
	}\right\}  .\label{frag-reg}
	\end{equation}

	\item For $xq_{\bot}\ll k_{\bot}\ll q_{\bot}$we have $D_B =D_A=k_{\bot}%
	^{2}$ and $D_C =q_{\bot}^{2}$,%
	\begin{align}
	\frac{dN}{d^{2}k_{\bot} dy}&\sim\alpha_{\text{s}}(x^{2}-2x+2)\left\{
	2C_{F}\frac{\,q_{\bot}^{2}x^{2}}{k_{\bot}^{4}}\right.\nonumber\\
	&\left.+C_{A}\frac{q_{\bot}^{2}%
	}{k_{\bot}^{2}}\left(  \frac{(1-x)^{2}+1}{q_{\bot}^{2}}-\frac{x^{2}}{k_{\bot
		}^{2}}\right)  \right\}  .  \label{cent-reg}
	\end{align}
	This formula does have a $1/k_{\bot}^{4}$ part, which is however suppressed in
	the large $N_{c}$ limit.
	
	\item Finally for $q_{\bot}\ll k_{\bot}$, D$_{A}=D_B=D_C =k_{\bot}^{2}$,%
	\begin{equation}
	\frac{dN}{d^{2}k_{\bot} dy}\sim\alpha_{\text{s}}(x^{2}-2x+2)\frac{q_{\bot}^{2}%
	}{k_{\bot}^{4}}\left\{  2C_{F}x^{2}+2C_{A}(1-x)\right\}  .\label{hi-rec}
	\end{equation}
	This formula is ``$x$-safe'' so that we can take both $x=0$ and $x=1$ limits and
	it agrees with the large $k_{\bot}$ limit of (\ref{xeq0}) and (\ref{xeq1}).
\end{itemize}

The reason why we need  
 to compare Eqs.~(\ref{xeq0}), (\ref{xeq1}), (\ref{cent-reg}) and (\ref{hi-rec}) only
to  the Bertsch-Gunion result of Ref.~\cite{gunion1982hadronization} is because we are interested 
in high transverse momentum for the gluon emission
where only the first order interactions in the strength of the sources 
 are
 required.  
A full treatment for the central region where such coherence effects are important is given {\it e.g.} in  
Ref.~\cite{Kovchegov:1998bi} and in Refs.~\cite{Kopeliovich:1998nw,Kopeliovich:1999am} where also the confinement
effects have been discussed. 
A treatment of the coherent region for the fragmentation region is included in \cite{kajantie2019gluon,Kajantie:2019nse}.

Unpacking these results a bit further, the condition $k_{\bot}\ll q_{\bot}$ means 
that we are looking at soft gluons in the sense that there is virtually no recoil of the emitting quark. 
Staying with the soft-gluon case, the condition $k_\bot\ll xq_\bot$ is equivalent to requiring that the emitted 
gluons have rapidity between zero and the final rapidity of the kicked quark -- we're essentially looking at the fragmentation 
region.   Eq.\,\eqref{frag-reg} then says: for recoil-less quarks, in the fragmentation region, the contribution from QED-like
 bremsstralhung falls off like $1/k_\bot^2$  while the BG contribution is constant in $k_\bot$.

The next case, $xq_\bot\ll k_\bot\ll q_\bot$, still considers  recoil-less qaurks but this time in the central region. 
Here the  QED-like bremsstralhung falls off rapidly as $1/k_\bot^4$ while the BG contribution is dominated 
by a $1/k_\bot^2$ fall-off, as seen in Eq.\,\eqref{cent-reg}.
In the final  case, $k_\bot\gg q_\bot$, we are looking at high recoil and 
Eq.\,\eqref{hi-rec} shows a $\sim 1/k_\bot^4$ fall-off in all regions. 
Here the fragmentation region corresponds to $x\geq 1/2$.


\section{Conclusion}
We have computed the contribution to gluon radiation of a particle scattering from the strong field of a nucleus.  As noted in Ref. \cite{Kajantie:2019nse}, the classical treatment of the particle computation breaks down in this region.  This result should allow a proper matching onto the high transverse momentum region of the emitted gluon, as in this region one can compute the radiation perturbatively, and
the contribution we present should be of leading order.  This paper therefore completes the determination of the  ingredients necessary to properly determine the initial conditions for matter produced in the fragmentation region of high energy heavy ion collisions.

\bigskip

\section*{Acknowledgements}

Larry McLerran wishes to gratefully acknowledge many useful discussions with Keijo Kajantie and Risto Paatelainen. Larry McLerran was supported, and  Mawande Lushozi, Michal Praszalowicz, Gongming Yu were partially supported by
the U.S. DOE under Grant No. DE-FG02-00ER41132.  Mawande Lushozi and Gongming Yu were partially supported under the Multifarious Mind grant provided by the Simons Foundation. Gongming Yu was partially supported by the National Natural Science Foundation of China under Grant No. 11847207, the International Postdoctoral Exchange Fellowship Program of China under Grant No. 20180010, and the China Postdoctoral Science Foundation Funded Project under Grant No. 2017M610663.




\begin{thebibliography}{}

\bibitem[Anishetty, Koehler and McLerran(1980)]{AKM1980fragmention}
R. Anishetty, P. Koehler and L. D. McLerran, Phys. Rev. D {22}, {2793} (1980).

\bibitem[McLerran(2018)]{M2018fragmention}
L. D. McLerran, EPJ Web Conf. {172}, {03003} (2018).

\bibitem[Gribov, Levin and Ryskin (1983)]{GLR1983Saturation}
L. V. Gribov, E. M. Levin and M. G. Ryskin, Phys. Rept. {100}, {1} (1983).

\bibitem[Mueller (1990)]{M1990Saturation}
A. H. Mueller, Nucl. Phys. B {335}, {115} (1990).

\bibitem[KLF (1977)]{KLF1977BFKL}
E.A. Kuraev, L.N. Lipatov and V.S. Fadin, Sov. Phys. JETP {45}, {199} (1977).

\bibitem[BL (1976)]{BL1976BFKL}
Ya. Ya. Balitsky and L. N. Lipatov, Sov. J. Nucl. Phys. {28}, {22} (1978).

\bibitem{mclerran1994computing}
L. D. McLerran and R. Venugopalan, Phys. Rev. D {49} , {2233}, (1994); Phys. Rev. D {49} , {3352} (1994).

\bibitem[Gelis and Venugopalan (2006)]{GV2006YangMills}
F. Gelis and R. Venugopalan, Acta Phys. Polon. B {37}, {3253} (2006).

\bibitem[Wong (1970)]{W1970YangMills}
S. K. Wong, Nuovo Cim. A {65}, {689} (1970).

\bibitem[Dumitru and McLerran(2002)]{dumitru2002protons}
A.~Dumitru and L.~D. McLerran, Nucl. Phys. A {700}, {492} (2002).
	
\bibitem[Kajantie \emph{et~al.}(2019)Kajantie, McLerran, and Paatelainen]{kajantie2019gluon}
K.~Kajantie, L.~D. McLerran, and R.~Paatelainen, Phys. Rev. D, {100}, {054011} (2019).

\bibitem{Kajantie:2019nse}
K.~Kajantie, L.~D.~McLerran and R.~Paatelainen,
Phys. Rev. D {101}, 054012 (2020).

\bibitem[Gelis, Stasto, and Venugopalan (2006)]{GSV2006ForawardLimit}
F. Gelis, A. M. Stasto and R. Venugopalan. Eur. Phys. J. C {48}, {489} (2006).

\bibitem[Iancu, Marquet, and Soyez (2006)]{IMS2006ForawardLimit}
E. Iancu, C. Marquet and G. Soyez, Nucl. Phys. A {780}, {52} (2006).

\bibitem[Gunion and Bertsch(1982)]{gunion1982hadronization}
J. F. Gunion and G. Bertsch, Phys. Rev. D {25}, {746} (1982).

\bibitem[Low (1975)]{L1975}
F. E. Low, Phys. Rev. D {12}, {163} (1975).

\bibitem[Nussinov (1976)]{N1976}
S. Nussinov, Phys. Rev. D {14}, {426} (1976).

\bibitem[KLF(1976)]{KLF1976LipatovVertex}
E. A. Kuraev, L. N. Lipatov and V.S. Fadin, Zh. Eksp. Teor. Fiz. {71}, {840} (1976)
[Sov. Phys. JETP {44}, {443} (1976)].

\bibitem{Kovchegov:1998bi}
Y.~V.~Kovchegov and A.~H.~Mueller,
Nucl. Phys. B \textbf{529}, 451-479 (1998).

\bibitem{Kopeliovich:1998nw}
B.~Z.~Kopeliovich, A.~V.~Tarasov and A.~Schafer,
Phys. Rev. C \textbf{59}, 1609-1619 (1999).

\bibitem{Kopeliovich:1999am}
B.~Z.~Kopeliovich, A.~Schafer and A.~V.~Tarasov,
Phys. Rev. D \textbf{62}, 054022 (2000)


\end{thebibliography}
\end{document}